\documentclass[pdftex]{article}
\usepackage{icrctc07}
\title{Calibration Techniques for VERITAS}
\shorttitle{VERITAS Calibration}
\authors{David Hanna$^{1}$ for the VERITAS Collaboration$^{2}$}
\shortauthors{D. Hanna et al}
\afiliations{$^1$Physics Department, McGill University, 3600 University Street, Montreal, QC H3A 2T8, Canada \\ $^2$For full author list see G. Maier ``VERITAS: Status and Latest Results'', these proceedings }
\email{hanna@physics.mcgill.ca}

\abstract{
VERITAS is an array of four identical telescopes designed 
for detecting and measuring astrophysical gamma rays with energies in excess of
100 GeV. Each telescope uses a 12 m diameter reflector to collect
Cherenkov light from air showers initiated by incident gamma rays and 
direct it onto a `camera' comprising 499 photomultiplier tubes read out
by flash ADCs. We describe here calibration methods used for determining the
values of the parameters which are necessary for converting the digitized PMT 
pulses to gamma-ray energies and directions. Use of laser pulses 
to determine and monitor PMT gains is discussed, 
as are measurements of the absolute throughput of the telescopes using 
muon rings.
}

\begin{document}
\maketitle
%Begin the section.

\section{Introduction}
Like all gamma-ray detectors which use the atmospheric Cherenkov technique, 
the VERITAS instrument is fundamentally quite simple. 
Each of its four telescopes consists of a 12 m reflector which directs 
Cherenkov light from air showers onto a matrix of 499 photomultiplier tubes 
(PMTs) which are read out using 500 MSample/s flash-analog-to-digital
converters (FADCs). In order to translate the digital information emerging
from the FADCs into a form which can be used to 
select gamma-initiated showers from background and determine the energy 
and direction of the incident gamma ray, one needs calibration constants.
These `constants' (which are not, strictly speaking, constant) need to be 
determined when commissioning the detector and monitored and adjusted 
periodically during the lifetime of the project.
In this paper we describe techniques employed by the VERITAS 
collaboration to accomplish this task; two techniques 
use a laser to determine the 
absolute gains of the PMTs and one uses Cherenkov images, generated by isolated
muons, for inter-telescope calibration and determination of absolute 
throughput.
Some of these issues are addressed independently with a remote LIDAR-like 
system described elsewhere in these proceedings \cite{Hui}. 

\section{The VERITAS Laser System}

For flat-fielding and gain monitoring, VERITAS uses a nitrogen laser ($\lambda$ = 337 nm, pulse energy 300 $\mu$J, pulse length 4 ns). 
The beam is sent through neutral density filters arranged in two sequential 
wheels, with 6 filters each, such that transmissions ranging from less than  
0.02\% to 100\% may be chosen.
It is then divided, approximately equally, among 10 optical fibres, four of 
which are routed to opal diffusers located on the optical axes of the 
telescopes, 4 metres from the PMTs in the cameras.
A fifth fibre supplies light to a PIN photodiode to provide a fast external 
trigger for FADC readout; self-triggers using only PMT information are also 
used for some applications. 
There is, at present, no independent monitor for measuring the pulse-to-pulse
fluctuations in the laser intensity (typically 10\%); these are monitored using
a sum over a large number of PMTs in each camera.

\subsection{Nightly Laser Runs}

A five-minute, 10 Hz laser run at nominal intensity is taken at the beginning
of each observing night.
The data obtained are used primarily for monitoring gain evolution and checking
for problems.
Other tests, described below, are done less frequently.
Since the opal diffuser spreads the laser light uniformly (to better than 1\%)
over the face of the
camera, the pulses can be used for flat-fielding the response of the channels.
The high voltages 
of the individual PMTs are adjusted so that the average pulse size in each 
channel is the same for all channels.
A PMT's average pulse size depends on the product of its 
photocathode's quantum efficiency and the efficiency for photoelectrons 
to be collected by its first dynode, as well as on the gain in the electron
multiplier stage. To a lesser extent it depends on the reflectivity of the 
Winston-cone light concentrator in front of each PMT.
The average pulse sizes are calculated and written to a database for use in 
off-line analysis.

The gain of the electron multiplier can be tracked separately using the 
daily laser data using the method of photostatistics. 
In this method, we remove laser fluctuations using 
a sum-over-PMTs monitor 
and the effects of electronics noise and night sky background are
measured in runs with zero laser intensity and unfolded. 
Then, 
to first order, we can state that the mean charge in a laser pulse is given 
by $\mu = G N_{pe}$ with $G$ the gain and $N_{pe}$ the mean number of 
photoelectrons arriving at the first dynode. 
Assuming that only Poisson fluctuations in $N_{pe}$ determine the width, 
$\sigma$, of the charge distribution, we have
$\sigma = G \sqrt{N_{pe}}$. 
Thus we can solve for gain as $G = \sigma^2/\mu$.
Taking into account statistics at the other dynodes, which are in general 
described by a Polya distribution, leads to a correction 
factor such that $G = \sigma^2/\mu/(1+\alpha^2)$ where $\alpha$ 
is the width parameter which would result from injecting only 
single photoelectrons into the dynode chain. 
For our PMTs and their associated dynode voltages we simulate $\alpha = 0.47$,
which results in a revised estimate for multiplier gain of 
$G' = 0.82~\sigma^2/\mu$.

\begin{figure}[h]
\vskip -1cm
\begin{center}
\includegraphics [width=0.8\textwidth]{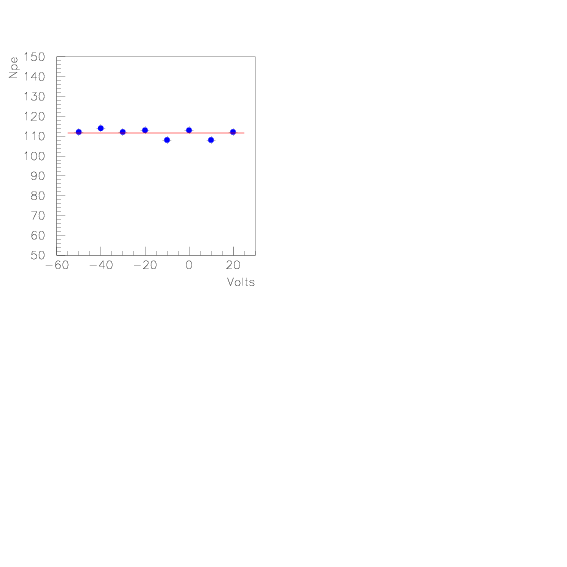}
\end{center}
\vskip -6.5cm
\caption{Mean number of photoelectrons captured by the first dynode of 
a representative PMT {\it vs} deviation from its nominal
high voltage setting. The flat line is to guide the eye and emphasize that
there is no significant change.}\label{hv_scan}
\end{figure}

As a check on this model, note that $\mu = G' N_{pe}$ or $N_{pe} = \mu/G'$.
This quantity is plotted for a representative PMT, in figure~\ref{hv_scan}, 
as a function of applied high voltages in steps from nominal HV. 
Except perhaps for an effect due to 
increased first dynode collection efficiency due to increased 
HV, we do not expect $N_{pe}$ to change and the plot shows that it is constant
over the range of voltages explored.

\subsection{Single Photoelectrons}

\begin{figure}[h]
\begin{center}
\includegraphics [width=0.35\textwidth]{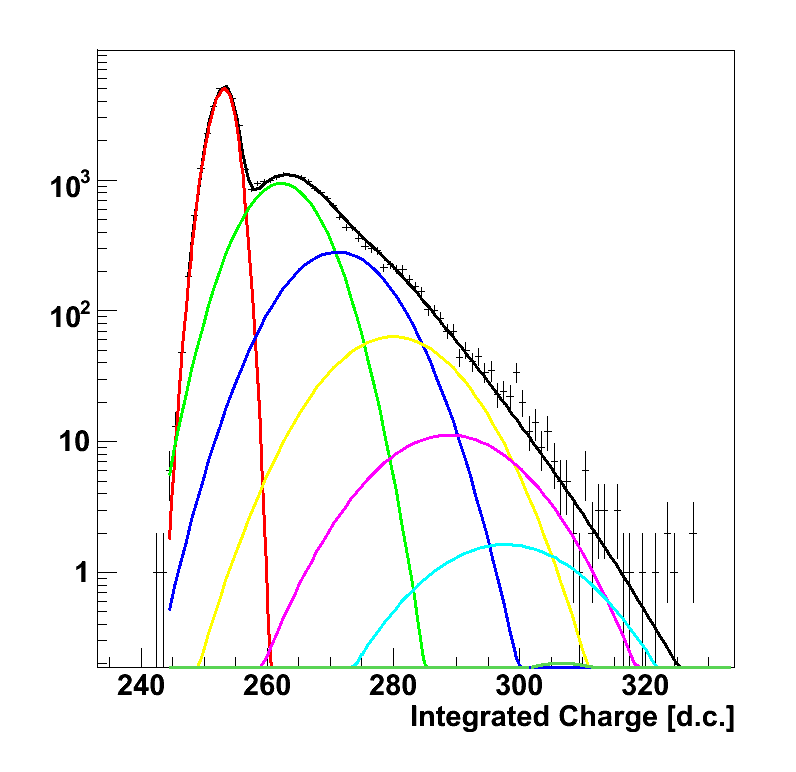}
\end{center}
\vskip -0.5cm
\caption{A pulse size spectrum made with highly attenuated laser pulses and
raised high-voltage.
The single photoelectron peak is clearly visible as the structure next to
the pedestal, which is the dominant feature. The data are fit with a 
sum of Gaussians as described in the text.}\label{spe}
\end{figure}

An alternative method for determining PMT gain is to directly measure the 
position of the single photoelectron peak in a pulse size spectrum. 
Again, this gives the gain of the electron multiplier structure (and any 
downstream electronics) and does not include effects of the photocathode.
To resolve the single photoelectron peak, we take special laser runs at very 
low intensity where the average number of photoelectrons resulting from each 
laser pulse is less than 1.0. 
The resulting spectrum consists of a pedestal, the single photoelectron peak, 
and small admixtures of two, three {\it etc} 
peaks with the relative sizes
of each component prescribed by Poisson statistics. 
We also rely on the constraint that the multi-photoelectron signals can be 
fit with the same parameters (mean and width) as the single photoelectron peak
(up to multiplicative factors). 
Allowing only a small number of free parameters results in robust 
fits to the spectra, an example of which is shown in figure~\ref{spe}.
In this example the relative width of the single photoelectron peak is found 
to be 0.48.
The average from a group of channels is 0.47.
The data for this study were obtained using twice the normal gain, in 
order to resolve more clearly the single photoelectron peaks. 
With such a gain the simulation predicts a relative width of 0.44.

In order to maintain good signal-to-noise for this measurement, 
we cover the camera with 
a thin aluminum plate with a 3 mm hole drilled at the location of the 
centre of each PMT.
This reduces the night sky background to the point where it is 
negligable compared with the laser light.
Indeed, with the telescope in stow position, one can perform single 
photoelectron laser runs in the presence of moonlight. 
This is an important consideration given that sufficient statistics 
($\sim$50000 shots) require nearly an hour of running.

\begin{figure}
\begin{center}
\includegraphics [width=0.35\textwidth]{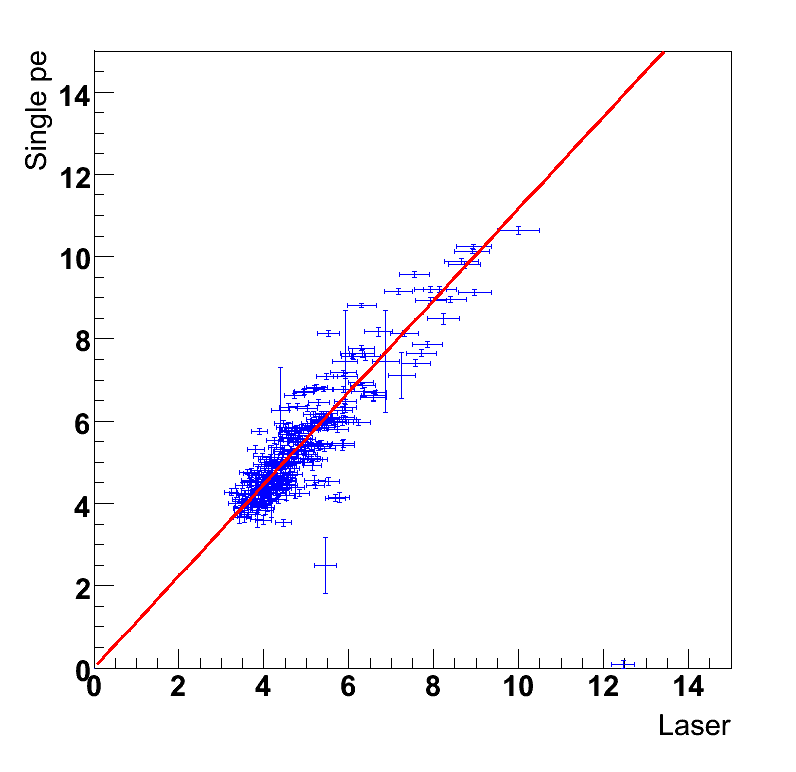}
\end{center}
\vskip -0.5cm
\caption{A comparison of gains determined using photostatistics (abscissa) 
with those determined from single photoelectron fitting (ordinate).
The slope of the correlation is approximately 1.1. 
It should be 1.0; the discrepancy is an indication of the present 
scale of the systematic error of the gain-measuring procedures.}
\label{spe_stat}
\end{figure}

The gain values 
which result from the method of photostatistics and from the single 
photoelectron fitting are in units of digital counts per photoelectron. 
A comparison of the two methods is shown in figure~\ref{spe_stat} where 
results from telescope 1 are shown.

The data points in this figure highlight the difference between the `multipler 
gain', which includes everything starting from the first dynode, and the 
`overall gain' which also includes the light concentrator cones and the 
photocathodes. Since the PMTs have all been flat-fielded according to the 
overall gain, the dispersion seen along the correlation line
in figure~\ref{spe_stat} is due to 
channel-to-channel differences in these `front end' components.

\section{Muon Rings}

Local muons are normally a nuisance for Cherenkov telescopes but they can be 
useful in providing a measurement of the optical throughput of the detector
\cite{Vacanti:1994pe,Humensky}.
Muons passing through the centre of the telescope with trajectories parallel 
to its optical axis will produce azimuthally symmetric rings in the camera.
The rings will have radii given by the Cherenkov angle of the muons (maximum 
value about 1.3 degrees) and the total number of photons expected 
in the ring can be calculated from the measured value of this angle.
Muons with non-zero impact parameters will produce arcs with an azimuthally 
dependent photon density and muons arriving at an angle
with respect to the telescope's axis will give rise to arcs with 
centres that are offset from the centre of the camera.

\begin{figure}
\begin{center}
\includegraphics [width=0.35\textwidth]{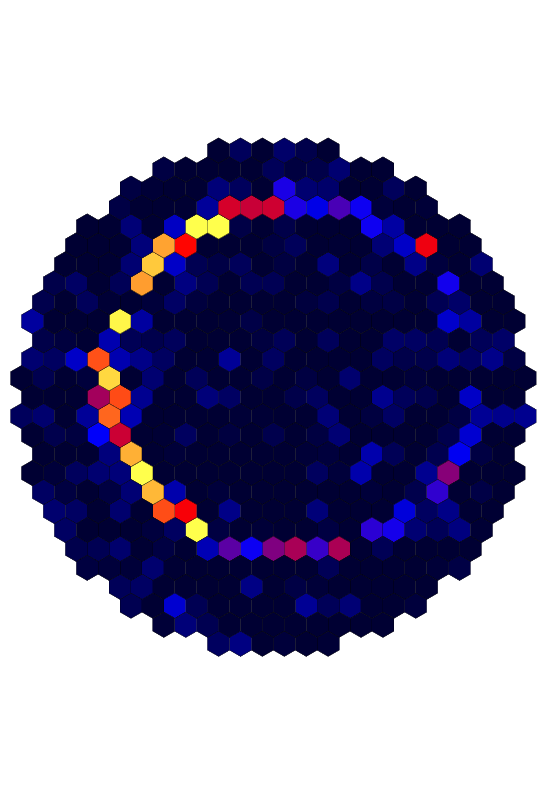}
\end{center}
\vskip -2cm
\caption{A muon image recorded by VERITAS telescope 1. 
Pulse size in each PMT is colour-coded - the azimuthal nonuniformity is the 
result of a non-zero impact parameter.}
\label{ring}
\end{figure}

Muon ring images can be obtained from normal data where they occur as part of 
hadronic showers.
The images are cleaned (channels are required to have a minimum pulse size and 
to be next to other channels with non-zero charge, otherwise they are set to 
zero) and a ring is fit to the image.
Further cleaning of the images, where charge deposits far from the fitted ring
are suppressed, removes light from other components of the shower
of which the muon was a member.
After this second cleaning the ring parameters are re-calculated. 
A cleaned image of a complete muon ring is shown in figure~\ref{ring}.

\begin{figure}
\vskip -1cm
\begin{center}
\includegraphics [width=0.9\textwidth]{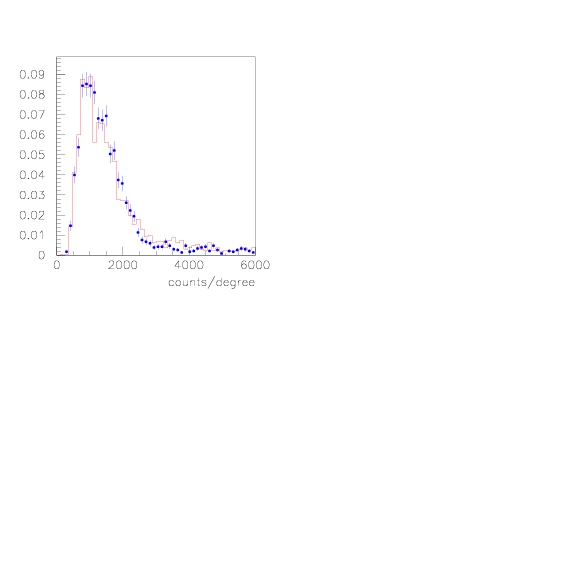}
\end{center}
\vskip -7.0cm
\caption{Detected charge in muon arcs, normalized to their 
lengths, for VERITAS telescopes 1 (histogram) and 2 (data points),
showing that they are well matched.}
\label{muon}
\end{figure}

Since the morphology and location of the muon ring allow the muon's trajectory 
to be calculated, it is possible to predict with precision 
the number of Cherenkov photons that should be collected by the camera. 
This requires knowing the reflectivity of the mirror facets, shadowing effects 
due to the camera support structure, {\it etc} so the detector response to 
local muons is a good check on our understanding of the instrument.
Absolute calculations are still in progress but certain relative measurements 
have already been implemented, such as inter-telescope calibration and 
month-to-month stability checks. 
An example is shown in figure~\ref{muon} where we histogram 
the summed charge in each muon arc, divided by its length, 
for two telescopes in the array. 
The overlap of the two histograms, normalized by the number of entries, 
shows that the telescopes are well balanced.

\section{Acknowledgements}

This research is supported by grants from the U.S. Department of Energy,
the U.S. National Science Foundation,
and the Smithsonian Institution, by NSERC in Canada, by PPARC in the UK and
by Science Foundation Ireland.

%\bibliography{libros}
\bibliography{icrc0702}
\bibliographystyle{plain}

\end{document}